\documentclass[useAMS,usenatbib]{mnras}
\usepackage{natbib}

\usepackage{graphicx}
\usepackage{epsfig}

\usepackage{txfonts}

\usepackage{hyperref}

\usepackage{float}
\usepackage{color}
\usepackage{lscape,graphicx}
\usepackage{journal_shortcuts}
\usepackage{amssymb}
\usepackage{multirow}
\usepackage{afterpage}
\usepackage{ulem}
\usepackage{threeparttable}
\usepackage{comment}
\usepackage{morefloats}
\usepackage{longtable}
\usepackage{pdflscape}
\usepackage{mdwlist}
\usepackage{upgreek}
\newcommand{\pb}{\hbox{Pa$\upbeta$}}
\newcommand{\hei}{\hbox{He~$\scriptstyle\rm I$}}
\newcommand{\caii}{\hbox{Ca~$\scriptstyle\rm II$}}

\title[The $M_{\rm{BH}} - \upsigma_\star$ of AGN2]{Detection of faint broad emission lines in type 2 AGN: III. On the $M_{\rm{BH}} - \upsigma_\star$ relation of type 2 AGN}

\begin{document}
\author[F. Ricci et al. ]{F. Ricci,$^{1,2}$\thanks{E-mail:riccif@fis.uniroma3.it} 
F. La Franca,$^{1}$
A. Marconi,$^{3,4}$ 
F. Onori,$^{5,6}$ 
F. Shankar,$^{7}$ 
R. Schneider,$^{8,9}$
\newauthor
E. Sani,$^{2}$
S. Bianchi,$^{1}$ 
A. Bongiorno,$^{9}$ 
M. Brusa,$^{10,11}$ 
F.~Fiore,$^{8,9}$
R. Maiolino,$^{12,13}$
\newauthor 
C. Vignali$^{10,11}$
\\
$^{1}$Dipartimento di Matematica e Fisica, Universit\`a Roma Tre,
   via della Vasca Navale 84, 00146 Roma, Italy\\
$^{2}$European Southern Observatory, Alonso de Cordova 3107, Casilla 19, Santiago 19001, Chile\\
$^{3}$Dipartimento di Fisica e Astronomia, Universit\`a degli Studi di Firenze, Via G. Sansone 1, 50019 Sesto Fiorentino, Italy\\
$^{4}$INAF - Osservatorio Astronomico di Arcetri, Largo E. Fermi 5, 50125 Firenze, Italy\\
$^{5}$SRON Netherlands Institute for Space Research, Sorbonnelaan 2, 3584 CA Utrecht, Netherlands\\
$^{6}$Department of Astrophysics/IMAPP, Radboud University, P.O. Box 9010, 6500 GL Nijmegen, the Netherlands\\
$^7$Department of Physics and Astronomy, University of Southampton, Highfield SO17 1BJ, UK\\
$^{8}$Dipartimento di Fisica, Sapienza Universit\`a di Roma, P.le Aldo Moro 2, I-00185 Roma, Italy\\
$^{9}$INAF - Osservatorio Astronomico di Roma, via Frascati 33, 00044 Monte Porzio Catone, Italy\\
$^{10}$Dipartimento di Fisica e Astronomia, Universit\`a di Bologna, viale Berti Pichat 6/2, 40127 Bologna, Italy\\
$^{11}$INAF - Osservatorio Astronomico di Bologna, via Ranzani 1, 40127 Bologna, Italy\\
$^{12}$Cavendish Laboratory, University of Cambridge, 19 J. J. Thomson Ave., Cambridge CB3 0HE, UK\\
$^{13}$Kavli Institute for Cosmology, University of Cambridge, Madingley Road, Cambridge CB3 0HA, UK
}

\date{Accepted 2017 June 18. Received 2017 June 18; in original form 2017 March 18}

\pagerange{\pageref{firstpage}--\pageref{lastpage}} \pubyear{2017}

\maketitle

\label{firstpage}

\begin{abstract}
Type 2 active galactic nuclei (AGN) represent the majority 
of the AGN population. 
However, due to the difficulties in measuring their black hole (BH) masses, 
it is still unknown whether they 
follow the same BH mass-host galaxy scaling relations valid for quiescent galaxies and type 1 AGN.
Here we present the 
locus of type 2 AGN having virial BH mass estimates in the $M_{\rm{BH}}-\upsigma_\star$ plane. 
Our analysis shows that the BH masses of type 2 AGN are $\sim0.9$ dex smaller than type 1 AGN at $\upsigma_\star\sim 185$ km s$^{-1}$, 
regardless of the (early/late) AGN host galaxy morphology.
Equivalently, type 2 AGN host galaxies have stellar velocity dispersions $\sim 0.2$ dex 
higher than type 1 AGN hosts at $M_{\rm BH}\sim10^7$ M$_\odot$.

\end{abstract}


\section{Introduction}
Supermassive black holes (SMBHs, $M_{\rm{BH}}>10^5$ M$_\odot$)
are believed to be ubiquitous, sitting at the centres of the spheroid 
of almost every galaxy.
Co-evolutionary models that 
link the growth of SMBHs and of their host galaxies 
are supported by the observation of 
tight scaling relations between the 
black hole (BH) mass $M_{\rm{BH}}$ and the
host bulge properties, e.g. bulge stellar velocity dispersion 
$\upsigma_\star$
\citep{ferrarese00, gebhardt00, marconihunt03, haringrix04,
gultekin09,
grahamscott13,mcconnellma13, kormendyho13, savorgnan15}; bulge luminosity
$L_{\rm bul}$ and mass $M_{\rm bul}$
\citep{dressler89, magorrian98, kormendyrichstone95, sani11}. 
All these relations are based on local samples of galaxies with dynamically measured 
BH masses. 
It is still debated whether or not these scaling relations 
should depend on the morphology of the 
galaxy 
(e.g. barred/unbarred \citealt{graham08}, 
early/late \citealt{mcconnellma13}, classical/pseudo-bulges 
\citealt{kormendyho13}). 
Furthermore, 
the galaxy samples used for calibration could suffer from a selection effect
due to the resolution of the 
BH gravitational sphere of influence, in favour of the more massive BHs \citep{shankar16, shankar17}. 

Active galactic nuclei (AGN) are thought to follow the 
same scaling relations observed in quiescent galaxies 
with dynamically measured $M_{\rm{BH}}$. 
In particular, reverberation mapped 
\citep[RM,][]{blandformcknee82, peterson93}
type 1 AGN (AGN1, where broad, FWHM $>1000$ km s$^{-1}$, 
optical emission lines are visible in their spectra) seem to reproduce the scaling relation $M_{\rm{BH}} - \upsigma_\star$ once 
the BH masses are scaled for the virial factor $f$ \citep{onken04, woo10, graham11, park12, grier13, hk14}, 
$M_{\rm{RM}}=f\times M_{\rm{vir}}$, where $M_{\rm{vir}}$ is 
the virial mass calculated from RM campaigns. 
Also low-mass BHs \citep{baldassare16} and narrow line Seyfert 1 galaxies
\citep[][but see \citealt{rakshit17}]{woo15}
 have $M_{\rm{BH}} - \upsigma_\star$ relations 
consistent with that found in quiescent galaxies.
However, recent works have found that AGN1 hosts 
	reside significantly below the $M_{\rm{BH}} - M_{\rm{bul}}$ 
	\citep{hk14} and $M_{\rm{BH}} - M_\star$ \citep{reines15} relations 
	of quiescent galaxies.
Furthermore, there is evidence that the BH-host scaling relations 
become less tight as soon as a broader range of BH masses \citep[e.g. $M_{\rm BH}<10^7$ M$_\odot$, see the megamaser sample of][]{greene16} and different galaxy morphologies \citep[i.e. disks and spirals,][]{kormendy11} are probed.

Although type 2 AGN (AGN2)\footnote{By saying AGN2 we refer to those X-ray selected AGN where there is no (Seyfert 2) or weak (intermediate 1.8 - 1.9) evidence of BLR, or even no lines at all \citep[X-ray Bright Optically Normal Galaxies, XBONG;][]{comastri02} in their optical spectra.} 
represent the majority of the AGN population \citep{LF05},
it is still unsettled whether they 
do follow the same scaling relations defined by quiescent galaxies. 
This is because $M_{\rm{BH}}$ are difficult to measure in type 2 AGN which 
lack optical broad emission lines. These virialized lines are used 
in single epoch virial estimators \citep{mj02, vp06} to directly measure BH masses. 
According to the unified AGN model \citep{antonucci93}, orientation is 
responsible for the 
appearance of broad emission lines coming from the broad line region (BLR) and hence for the AGN1/AGN2 classification,
due to the presence of a dusty structure along the line of sight
\citep[e.g. the torus, but see e.g.][]{mezcua16}. 
In this scenario, AGN1 and AGN2 are expected to follow 
the same scaling relations, and therefore BH masses of 
AGN2 are usually estimated with (indirect) proxies, such as $\upsigma_\star$, 
$L_{\rm bul}$ and $M_{\rm bul}$.
These indirect estimates are often also used to 
evaluate the AGN BH mass function \citep[BHMF; e.g.][]{heckman04}. 

In order to directly measure the BH masses of AGN2, 
we have calibrated virial relations based on 
the hard X-ray luminosity and on the width of 
the most relevant near-infrared (NIR; 0.8-2.5 $\mu$m)
and optical emission lines \citep{LF15, LF16, ricci17}. 

\begin{figure}
	\centering
	\vspace{-0.15in}
	\includegraphics[width=0.9\columnwidth]{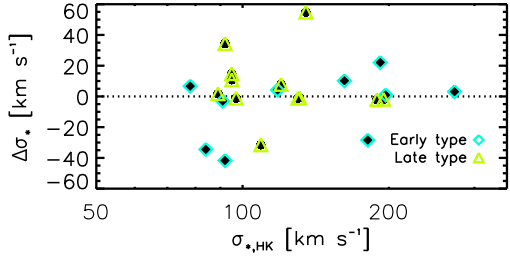}
	
	\caption{Difference between the HyperLeda stellar velocity dispersion and 
		the one reported by \citet{hk14}, as a function of 
		the latter. The dotted line marks the zero offset.
		Galaxies are plotted according to the host morphology
		(early type as cyan diamonds, late type as green triangles).}
	\label{Fig:ss}%
\end{figure} 

We have then carried out a
systematic search 
to detect faint virialized broad line components in the 
NIR spectra 
of hard X-ray selected obscured and intermediate
class AGN. 
We have observed the NIR spectra of 41 AGN2
at redshift $z \leq 0.1$, randomly drawn from the {\it Swift}/BAT 70-month
catalogue \citep{baumgartner13}. 
Data reduction, spectral analysis and line
fitting parameters  
have been published in the first paper of the series \citep[Paper I]{onori17a}. 
Broad virialized components in 
the NIR emission lines (i.e. \pb\ and \hei\ $\uplambda$ 10830 \AA)
have been measured in 13 out of 41 ($\sim$30\%)
of the selected AGN2. 
This starting sample  
has been extended with 
four AGN2 included in the
{\it Swift}/BAT 70-month catalogue whose 
FWHM NIR lines, or spectra, were already published.  
These 17 AGN2 have been used in a companion paper  
\citep[Paper II]{onori17b} to derive 
the first direct virial $M_{\rm{BH}}$ in AGN2.
The AGN2 virial $M_{\rm{BH}}$ have been computed 
using the virial estimators calibrated in \citet{ricci17}
which are based on the broad NIR FWHM and on the hard X-ray 
14-195 keV luminosity.
We found that 
when comparing AGN1 and AGN2 of the 
same X-ray luminosity, $\log L_{14-195\, \rm{keV}}\sim43.5$ erg s$^{-1}$,
the average FWHM of
the BLR in AGN2 is $\sim$0.25 dex smaller than measured in the control sample of
RM AGN1. As a consequence, AGN2 have 0.5 dex smaller $M_{\rm{BH}}$ and 
higher Eddington ratios than RM AGN1
with the same intrinsic X-ray luminosity.
These findings do not support a ''pure'' orientation-based unified model,
possibly indicating that AGN2 could 
be associated with 
an earlier evolutionary stage
or may comprise different physical configurations or mechanisms for BH growth.
 
In this paper, we present for the first time the local
$M_{\rm{BH}}-\upsigma_\star$ plane for AGN2 with virial BH masses in order to understand whether 
AGN2 share the same properties of AGN1. 

Throughout the paper we assume a flat $\Lambda$CDM cosmology
with cosmological parameters $\Omega_\Lambda$ = 0.7, $\Omega_M$ = 0.3 and
$H_0$ = 70 km s$^{-1}$ Mpc$^{-1}$. Unless otherwise stated, all the quoted
uncertainties are at 68\% (1$\upsigma$) confidence level.

\begin{table*}
	\begin{minipage}{180mm}
		\caption{General properties of the AGN2 sample. 	Columns are: 
			(1) galaxy name; 
			(2) redshift from NED;  
			(3) activity type; 
			(4) logarithm of $M_{\rm{BH}}$, computed with the solution a3 of Table 4 from \citet[][based on optical-NIR broad lines and $L_{14-195 \rm{keV}}$ luminosity]{ricci17}; uncertainties on $M_{\rm{BH}}$ are only statistical, i.e. do not take into account the population spread ($\sim$ 0.5 dex);
			(5) reference of the BH mass;
			(6) bulge stellar velocity dispersions $\upsigma_\star$;
			(7) reference for $\upsigma_\star$, L17=\citet{lamperti17}, HL=HyperLeda;
			(8) morphological type retrieved from HyperLeda. 
			$^\dag$Classified after visual inspection 
			of a DSS blue image.}
		\label{Tab:1}
		\begin{center}                          
			\begin{tabular}{l c l c c c c l}
				\hline 
				\noalign{\smallskip}
				Galaxy & $z$ & activity & $\log M_{\rm{BH}}$ & ref $M_{\rm{BH}}$ & $\upsigma_\star$ & ref $\upsigma_\star$  & morphological \\
				& & type &[M$_\odot$] & & [km s$^{-1}$] & & type \\
				(1) & (2) & (3) & (4) & (5) & (6) & (7) & (8)\\
				\noalign{\smallskip}
				\hline
				\noalign{\smallskip}
				
				2MASX J07595347+2323241 & 	0.0292	&	2		& 7.78  $\pm$   0.09	&This work &	230 $\pm$ 21 & L17 &Late \\
				2MASX J18305065+0928414	&	0.0190	&	2		& 7.04  $\pm$   0.09	 &Paper II &	196 $\pm$ 19 &This work	&Late$^\dag$	\\	
				ESO 234-G050			&	0.0088	&	2		& 6.00  $\pm$   0.10	 &Paper II &	69  $\pm$ 1	 &This work	&Early	\\			
				MCG -05-23-016			&	0.0850	&	2		& 7.22  $\pm$   0.06	 &Paper II &	172 $\pm$ 20 &HL	&Early	\\		
				Mrk 348  				&	0.0150	&	2/FSRQ	& 7.23  $\pm$   0.08	 &Paper II &	141 $\pm$ 29 &HL	&Early 	\\		
				NGC 1052 				&	0.0050	&	2		& 6.63  $\pm$   0.09	 &Paper II &	209 $\pm$ 4\phantom{a}	 &HL	&Early	\\			
				NGC 1275 				&	0.0176	&	2		& 7.46  $\pm$   0.06	 &Paper II &	243 $\pm$ 13 &HL	&Early	\\			
				NGC 1365 				&	0.0055	&	1.8		& 6.65  $\pm$   0.09	 &Paper II &	143 $\pm$ 19 &HL	&Late	\\		
				NGC 2992 				&	0.0077	&	2		& 6.72  $\pm$   0.08	 &Paper II &	160 $\pm$ 17 &HL	&Late	\\		
				NGC 4395 				&	0.0013	&	1.9		& 5.14  $\pm$   0.07	 &Paper II &	27  $\pm$ 5	 &HL	&Late	\\		
				NGC 5506				&	0.0062	&	1.9		& 6.86  $\pm$   0.09	 &This work&	174 $\pm$ 19 &HL	&Late	\\		
				NGC 6221 				&	0.0050	&	2		& 6.46  $\pm$   0.06	 &Paper II &	64  $\pm$ 2	 &This work	&Late	\\		
				NGC 7465 				&	0.0065	&	2		& 6.54  $\pm$   0.10	 &Paper II &	95  $\pm$ 4	 &HL	&Early	\\

				\noalign{\smallskip}
				\hline
				
			\end{tabular}
		\end{center} 
	\end{minipage}
	\normalsize
\end{table*}

\section{Data}\label{sec:data}
\subsection{Sample selection}
Stellar velocity dispersion measurements $\upsigma_\star$ 
are available on HyperLeda\footnote{\url{http://leda.univ-lyon1.fr/}} 
for 8 out of the 17 AGN2 with NIR broad lines presented in Paper II.
Using the optical long slit spectra published in Paper I we have measured 
$\upsigma_\star$ for three additional AGN2.
We fitted with Gaussian profiles the \caii\ triplet $\uplambda$ 8500.36, 8544.44, 8664.52 \AA\ 
absorption lines extracted with a 1$''$ aperture, corresponding to 397, 192, and 92 pc for 2MASX J18305065+0928414, ESO 234-G050, and NGC 6221, respectively.
The HyperLeda database \citep{paturel03} presents
measurements which have been homogenized to a common aperture $r_{\rm{HL}}=0.595$ kpc.
In order to convert our $\upsigma_\star$ measurements 
to the common radius $r_{\rm{HL}}$
we corrected for the aperture effects using 
the relation $\upsigma_\star (r) \propto r^{-0.066}$
of \citet{cappellari06}.

We added to this sample 
the sources NGC 5506 and 2MASX J07595347+2323241 that have broad \pb\ lines measured by \citet{lamperti17}. 
For NGC 5506 a $\upsigma_\star$ measurement is available on HyperLeda.
For 2MASX J07595347+2323241, \citet{lamperti17} fitted
the CO 
band-heads in the H-band (1.570 - 1.720 $\mu$m)
from a long slit spectrum (aperture 0.8$''$, corresponding
to 480 pc).
As done previously,
$\upsigma_\star$ was corrected for aperture effects. 
Table \ref{Tab:1} lists the general properties of the final sample of 13 AGN2 with measured $\upsigma_\star$ and virial BH masses, which have all been calculated with solution a3 of Table 4 of \citet{ricci17} assuming an average virial factor $\langle f \rangle=4.31$.
This virial factor has been derived by \citet{grier13} 
by requiring that RM AGN1 reproduce the $M_{\rm BH}-\upsigma_\star$ relation
found in quiescent galaxies by \citet{woo13}.
In all the following analyses, we excluded 
the most deviating late-type AGN2, NGC 4395 which is 
one of the least massive active BH known.
It is a Sd bulge-less galaxy whose 
stellar velocity dispersion seems to be rotation-dominated 
also in the inner-part of the host galaxy \citep{denbrok15}.

As control sample of AGN1 we adopted 
the RM sample of 43 AGN1
presented in \citet{hk14}, who list $\sim$90\% of the RM sample along with available bulge morphology, classified as elliptical, classical, or pseudo-bulges.
Reliable measurements of central stellar velocity dispersion 
are available for a total of 31 sources. 
We considered also one additional RM AGN1, Fairall 9, 
classified as a classical bulge by \citet{hk14},
whose stellar velocity dispersion is 
available on HyperLeda.
The final control sample of RM AGN1 lists 32 sources.
The BH masses adopted for RM AGN1 are 
$M_{\rm{BH}}=f\times M_{\rm{vir}}$, where
$M_{\rm{vir}}$ are the updated virial 
masses also used in the calibrating sample by \citet{ricci17} 
and $\langle f \rangle = 4.31$.

\subsection{Stellar velocity dispersion measurements}

As described in the previous section, 
the AGN2 stellar velocity dispersions
were retrieved from HyperLeda and are extracted at 0.595 kpc,
while the stellar velocity dispersions of the 
RM AGN1 sample are calculated at the effective radius $R_e$.
In order to evaluate whether 
there are systematic offsets among the two databases, 
we compare 
the HyperLeda stellar velocity dispersion $\upsigma_{\star,\rm{HL}}$
and the stellar velocity dispersion measurements 
collected by \citet{hk14}, $\upsigma_{\star,\rm{HK}}$, for the sample of 22 RM AGN1
that have both.
Figure \ref{Fig:ss}
shows the offset between the two 
measurements
$\Delta \upsigma_\star = \upsigma_{\star,\rm{HL}} - \upsigma_{\star,\rm{HK}} $ 
as a function of $\upsigma_{\star,\rm{HK}}$.
The dotted line in Figure \ref{Fig:ss} marks the zero offset.
The majority of the sources shows negligible offsets 
$|\Delta \upsigma_\star| < 10$ km s$^{-1}$, and
the average is consistent with zero ($1\pm4$ km s$^{-1}$).

We divided RM AGN1 in early and late-type galaxies, 
as shown in Figure \ref{Fig:ss} (cyan diamonds are early-type and green triangles 
are late-type galaxies).
Indeed it is known that in late-type galaxies with a rotating stellar disk, the 
line-of-sight velocity dispersion could be broadened due to
the disk rotation \citep{bennert11, harris12, kang13}.
However, as shown in Figure \ref{Fig:ss}, this seems not to be the case for RM AGN1
because the most deviating measurements 
$|\Delta \upsigma_\star| > 30$ km s$^{-1}$ are equally late- and early-type AGN1.
The average offset in both early-type ($7\pm6$ km s$^{-1}$) and 
late-type ($-6 \pm 7$ km s$^{-1}$) are again almost consistent with 
zero.

We also checked whether 
in our sample of AGN2
the HyperLeda values have been extracted at $r$ significantly larger than
$R_e$. Indeed the disk rotational broadening in late-type galaxies 
should be higher outside the spheroid. 
\citet{oohama09} reported that the average $R_e$ in SDSS late-type 
galaxies (Sa, Sb, Sc) is $\sim 2.7$ kpc, hence the $\upsigma_\star$ 
retrieved from HyperLeda ($r=$0.595 kpc) should not contain substantial
rotational velocity contamination since the extraction is located
well inside this average value.

\begin{figure}
	\hspace{-0.5in}{\includegraphics[scale=0.52]{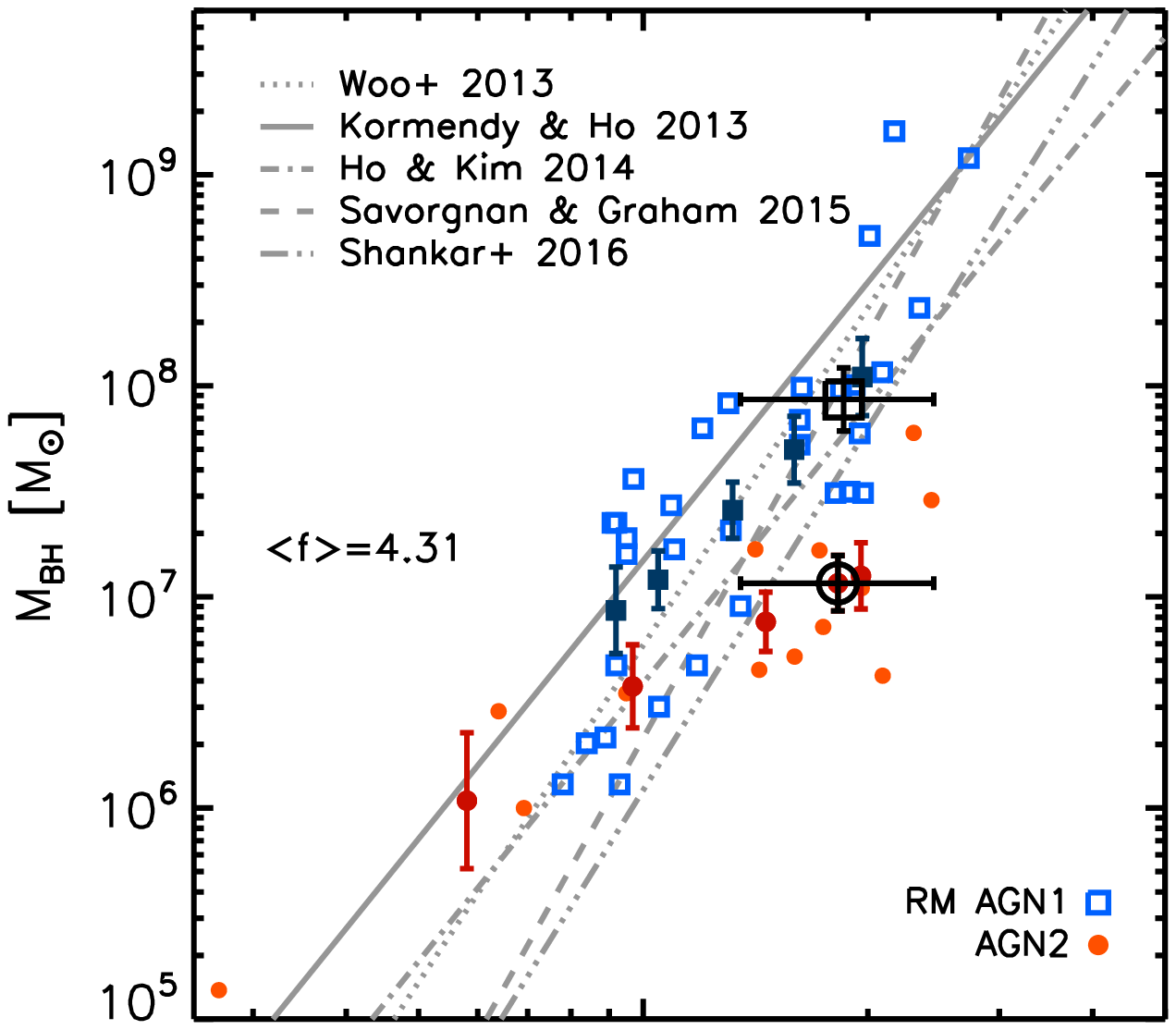}} 
	\hspace{-0.92in}{\includegraphics[scale=0.52]{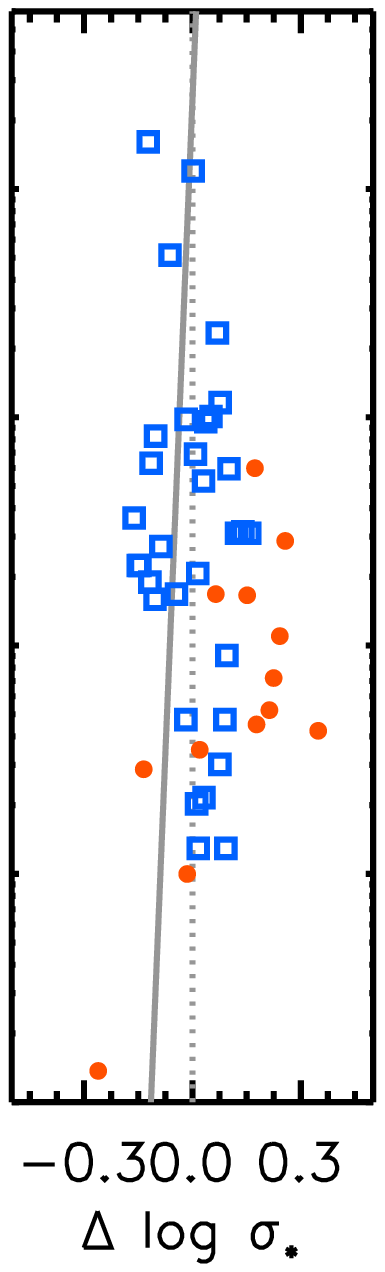}} \vspace{-0.79in}
	
	\hspace{-0.5in}{\includegraphics[scale=0.52]{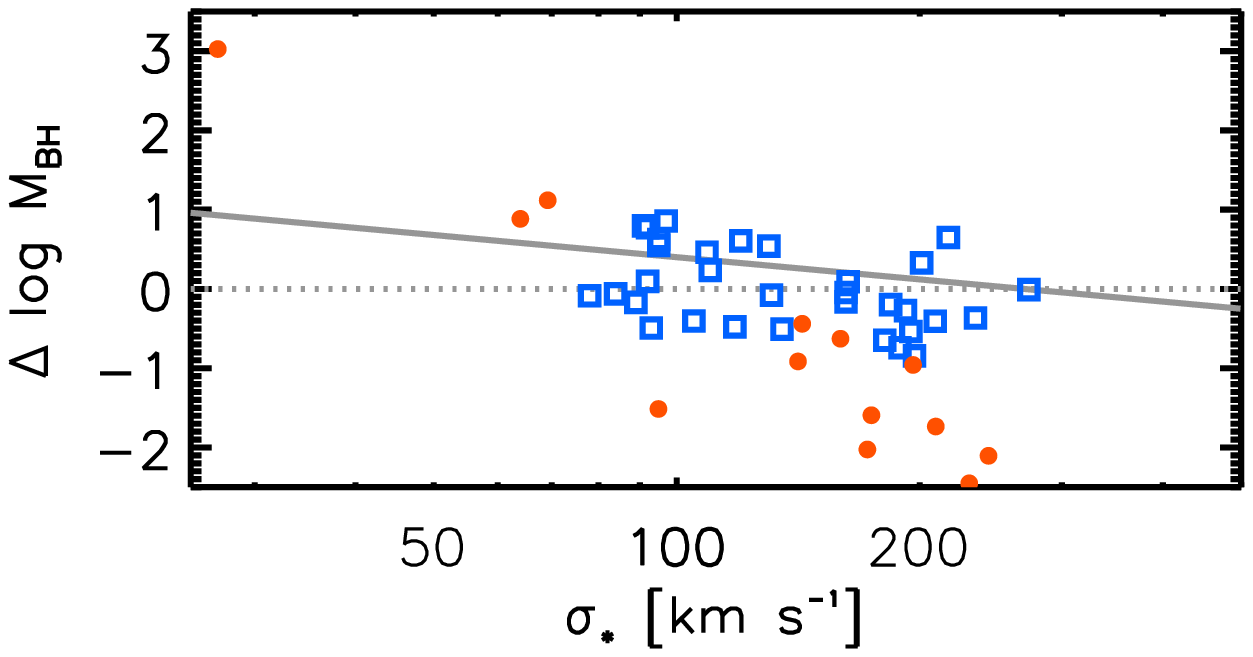}}
	
	\caption{The $M_{\rm{BH}} - \upsigma_\star$ plane for local samples of 
		RM AGN1 (blue open squares) and AGN2 (red filled circles), together with
		average $M_{\rm{BH}}$ of the RM AGN1 (dark blue) and AGN2 (dark red),
		computed in (not independent) bins of stellar velocity dispersion.
		The black open square (circle) shows the $M_{\rm{BH}}$ average value of the RM AGN1 (AGN2) sample in the 
		$135< \upsigma_\star <250$ km s$^{-1}$ stellar velocity 
		bin and has been plotted at the position of the average $\upsigma_\star$.
		Some relations from literature are also reported (see text for details).
		{\it Bottom:} residuals of $M_{\rm{BH}}$ with respect to the  
		relations of \citet[][dotted grey line]{woo13} and \citet[][solid line]{kormendyho13}.
		{\it Right:} same as bottom panel but for $\upsigma_\star$.}
	\label{Fig:Msbin}
\end{figure}

\begin{figure}
	\hspace{-0.5in}{\includegraphics[scale=0.52]{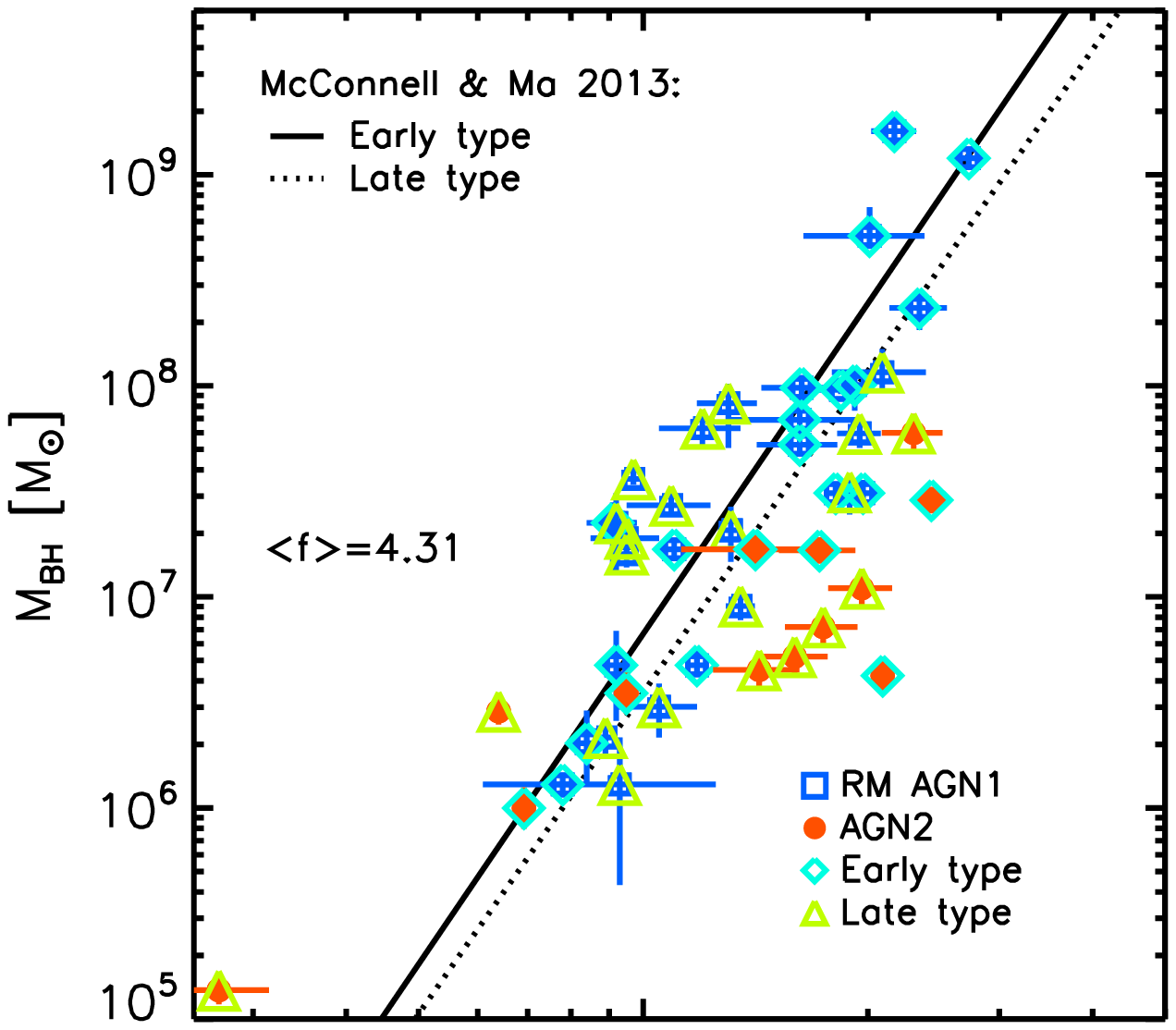}} 
	\hspace{-0.92in}{\includegraphics[scale=0.52]{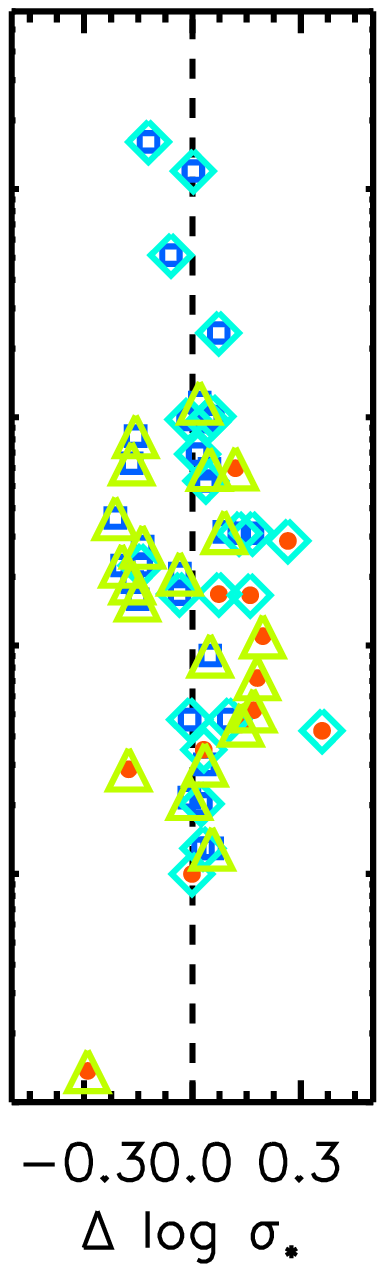}} \vspace{-0.79in}
	
	\hspace{-0.5in}{\includegraphics[scale=0.52]{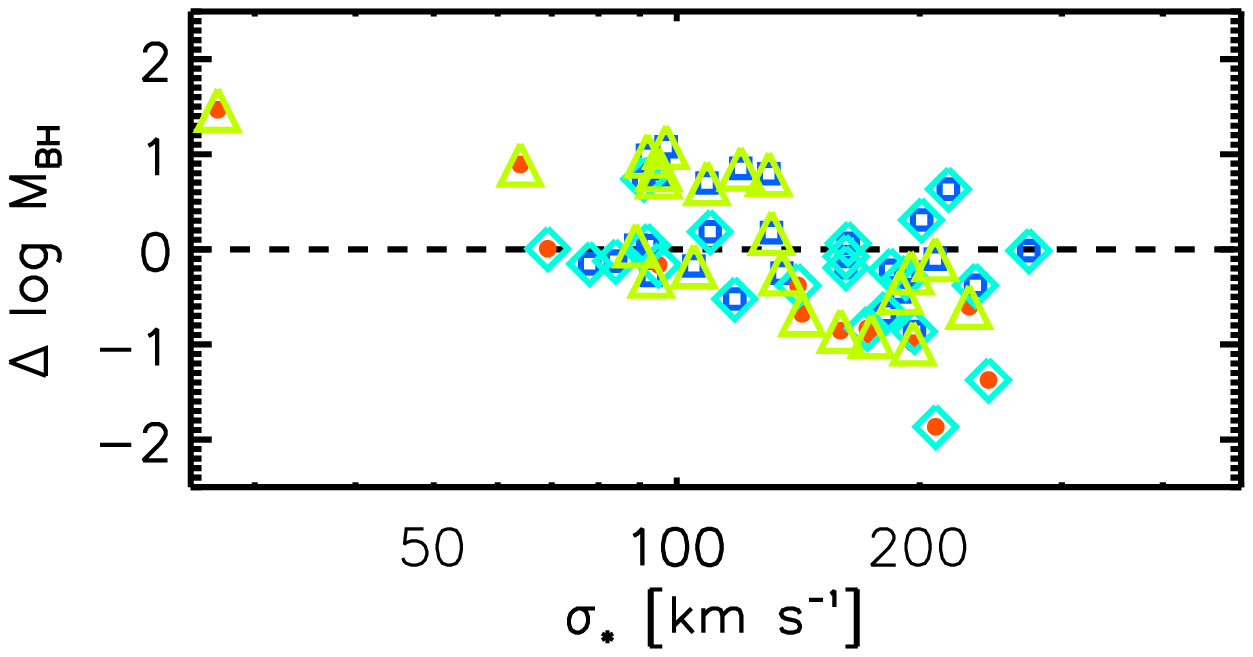}}
	\caption{The $M_{\rm{BH}} - \upsigma_\star$ plane for local samples of 
		RM AGN1 (blue) and AGN2 (red).
		Galaxies are further divided into early-type (cyan diamonds)
		and late-type (green triangles). The $M_{\rm{BH}} - \upsigma_\star$ calibrated by 
		\citet{mcconnellma13} for early-type (solid line) and late type (dotted line) are also shown.
		{\it Bottom:} residuals of $M_{\rm{BH}}$ versus the $M_{\rm{BH}}$ expected from the 
		relations of \citet{mcconnellma13}. 
		Residuals are calculated separately for late and early type using the pertaining scaling relations.
		{\it Right:} same as bottom panel but for $\upsigma_\star$ versus $M_{\rm{BH}}$.}	
	\label{Fig:MsEL}
\end{figure}

\section{Results}
Figure \ref{Fig:Msbin} shows 
the $M_{\rm{BH}} - \upsigma_\star$ plane for local samples of 
	RM AGN1 (blue open squares) and AGN2 (red filled circles), together with
the average BH masses of RM AGN1 (dark blue filled squares)
and AGN2 (dark red filled circles) computed in 
logarithmic bins of 
stellar velocity dispersion. 
Average $M_{\rm{BH}}$ are plotted at the average $\log \upsigma_\star$
of the AGN within each velocity dispersion bin.
Given the relatively small size of the samples, qualitative trends 
are better seen if the data are shown in not independent, 0.6 dex wide 
logarithmic $\upsigma_\star$ bin.
The black open square (circle) in Figure \ref{Fig:Msbin} shows the resulting average $M_{\rm{BH}}$ value of the RM AGN1 (AGN2) sample in the $135<\upsigma_\star<250$ km s$^{-1}$ stellar velocity bin where most of the two populations overlap and has been plotted at the position of the average $\upsigma_\star$.
The average BH masses in this bin are:
$\log (M_{\rm{BH}} / {\rm M}_\odot)= 7.06\pm0.13$ for AGN2 and 
$\log (M_{\rm{BH}} / {\rm M}_\odot) = 7.93\pm0.15$ for RM AGN1.
Hence at fixed $\upsigma_\star \simeq 184$ km s$^{-1}$, 
BH masses of AGN2 are $0.87$ dex smaller than in RM AGN1,
even though 
the same virial $f$-factor has been adopted in their derivation. 
Equivalently, 
in the overlapping BH mass bin $4\times 10^6 <M_{\rm BH}<2\times 10^7$ M$_\odot$,
AGN2 show higher stellar velocity dispersions: 
$\upsigma_\star=169\pm10$ km s$^{-1}$ for AGN2 and
$\upsigma_\star=106\pm 7$ km s$^{-1}$ for RM AGN1.
This means that at 
fixed $M_{\rm BH}\simeq 10^7$ M$_\odot$, the 
stellar velocity dispersion in AGN2 hosts is $\sim$0.2 dex higher than RM AGN1.
The bottom and right panels of Figure \ref{Fig:Msbin}
report the residuals in BH masses $\Delta \log M_{\rm{BH}}$ and stellar velocity 
dispersions $\Delta \log \upsigma_\star$. 
The BH masses residuals are computed as 
the logarithmic difference between the virially measured $M_{\rm{BH}}$ 
and that expected from the
scaling relation of \citet{woo13}.
For comparison, the residual of the relation 
of \citet[][solid grey line]{kormendyho13} is also reported.
As the relation of \citet{woo13}
is the reference for the calibration of the $f$-factor 4.31
that we are using,
the average residual in BH mass ($0.00\pm0.09$) and stellar velocity dispersions ($0.00\pm0.02$)
in the RM AGN1 are, as expected, consistent with zero.
On the contrary, the average 
residual for AGN2 is
$\Delta \log M_{\rm{BH}}=-0.99\pm0.37$ 
or equivalently 
$\Delta \log \upsigma_\star = 0.14\pm0.04$. 

If RM AGN1 are compared to the relation from \citet{kormendyho13},
the resulting average residuals are $\Delta \log M_{\rm{BH}} = -0.27\pm0.08$ 
and $\Delta \log \upsigma_\star = 0.06\pm0.02$.
These differences roughly correspond to 
changing the average $f$-factor from 4.31 to 6.2,
which is the value needed for RM AGN1 (calibrated by \citealt{hk14})
to reproduce the relation found in quiescent galaxies by \citet{kormendyho13}.
The same analysis applied to AGN2 confirms that
AGN2 show smaller BH masses at fixed $\upsigma_\star$, 
having average $\Delta \log M_{\rm{BH}}=-1.01\pm0.16$ (and $\Delta \log \upsigma_\star=0.23\pm0.04$).  
Thus our analysis suggests that AGN2 have smaller BH masses 
	than AGN1 at fixed $\upsigma_\star$ or equivalently 
	that the AGN2 host galaxies have higher bulge stellar velocity dispersions 
	at fixed $M_{\rm{BH}}$.

We have investigated whether these differences can be attributed 
to the host galaxy morphologies.
The central panel of Figure \ref{Fig:MsEL} shows the 
distribution of AGN2 (red) and RM AGN1 (blue) in the
$M_{\rm{BH}} - \upsigma_\star$ plane,
where the host galaxies are
divided into early- (cyan diamonds) and late-type (green triangles).
AGN2 hosts are divided into 6 early-type and 6 late-type, while among the RM AGN1 17 are early-type and 15 are late-type galaxies.
The scaling relations derived by \citet{mcconnellma13} separately 
for early-type (solid black line) and late-type (dotted black line) quiescent galaxies 
are also shown for comparison.
While the RM AGN1 are distributed around each of the two scaling relations obtained from quiescent galaxies,
the AGN2 lie below them, independently of the AGN host morphology.
The BH masses (stellar velocities) residuals are computed as 
the logarithmic difference between the virially measured $M_{\rm{BH}}$ ($\upsigma_\star$)
and the values expected for the same morphological class using the correlations by \citet{mcconnellma13}.
Residuals are reported in the bottom and right panels of Figure \ref{Fig:MsEL}.
The average $\Delta \log M_{\rm{BH}}$ of early-type (late-type)
RM AGN1 is $-0.09\pm0.10$ ($0.32\pm0.14$), while 
for early-type (late-type) AGN2 is $-0.77\pm0.30$ ($-0.52\pm0.29$). 
This analysis confirms that AGN2 
have lower BH masses than expected, regardless of the 
host galaxy morphology. 
For the late-type sample of RM AGN1 
the $\Delta \log M_{\rm{BH}}$ is 
not consistent with zero.
However, these residuals 
are dependent on the choice of the 
scaling relation used for comparison.
The relation for late-type galaxies from \citet{mcconnellma13},
which is rather steep (slope of $\sim 5$),
systematically under-predicts the BH masses 
at low stellar velocity dispersion ($\upsigma_\star\lesssim150$ km s$^{-1}$),
where most of the late-type RM AGN1 are located.

\section{Discussion}
According to our analysis, 
AGN2 have smaller BH masses than RM AGN1 at fixed bulge stellar velocity dispersion or, alternatively, 
larger velocity dispersions at fixed BH masses.
This result nicely complements
our previous findings in Paper II,
where we showed that AGN2 have smaller BH masses than RM AGN1
at fixed intrinsic hard X-ray luminosity.
The AGN2 BH masses have been derived using the 
virial estimator calibrated by \citet{ricci17},
that is based on the measurement of the NIR broad 
FWHM and on the hard X-ray 14-195 keV luminosity.
These quantities are most suitable 
for estimating the virial BH masses of low-luminosity AGN1 and in AGN2 because 
both the NIR and X-ray bands
are less dependent on 
obscuration or reddening than the optical. 

In Paper I we analysed possible biases on the 
detection of NIR broad emission lines in our AGN2 sample. No connection 
was found between the NIR broad line detectability and NIR flux, X-ray flux and 
luminosity, EW, FWHM, S/N, galaxy orientation and $N_H$ as measured in the 
X-rays. No BLR was found for the most (heavily, $\log N_H > 23.7$ cm$^{-2}$) 
X-ray obscured sources. In Paper II we further tested possible biases 
on the subsample of AGN2 that showed NIR virialized broad lines. 
Our previous analysis suggests that the NIR FWHMs are not probing 
only the outer (slower rotating) part of the BLR as we do not find any evidence of correlation 
between the FWHM and obscuration ($N_H$) or extinction ($A_V$). 
Therefore the subsample of AGN2 with NIR broad lines has no clear difference with the 
AGN2 for which we did not find NIR broad lines, and thus 
our sample of AGN2 with virial $M_{\rm BH}$ could be considered as a representative 
sample of Compton thin X-ray selected type 2 AGN.
There could be several reasons why we did not detect NIR broad lines 
in the whole sample of AGN2: i) they could be ``true'' Seyfert 2, i.e. truly lack a BLR \citep{tran01, elitzur09}; 
ii) as the AGN emission is variable, it could be that we observed the source in a low state;
iii) in clumpy torus models, 
the type 1-2 classification is probabilistic and depends on whether 
	there or not there is a clump along the line of sight, and so stochastic 
	variations in the dust distribution in any given source could be a reason 
	for non-detection;
iv) in particular for the more obscured sources
it could be that 
the NIR did not allow us to completely
penetrate the strong obscuration 
of the central engine.
Regardless of the possible biases of the 
NIR detectability, it seems unlikely that the 
missed NIR BLR are hosted in AGN2 having 
significantly different $M_{\rm BH}$ than the 
measured ones.
A possible observational strategy to supersede or test these hypotheses 
is moving to longer wavelengths that would be even less affected 
by dust absorption. 
Also, the detection of broad lines would benefit of  
IFU observations with high spatial resolution. 
Alternatively, our NIR campaign could be complemented by 
spectropolarimetric observations. However, 
spectropolarimetry requires high S/N to detect the low linear polarization signal 
typically observed in AGN \citep[$\sim$1-5\%;][]{antonuccimiller85}. 
While spectropolarimetry relies on the presence of a scattering region with sufficient 
covering factor and optical depth to allow scattered light to be observed and 
provide a ``periscope'' view of the inner part of the AGN, infrared spectroscopy 
offers a direct view of the BLR, as soon as it penetrates the dust.

In this letter we discussed how the difference in BH masses can not be ascribed to 
biases in the measurement of the bulge stellar velocity dispersion due to rotation in the host galaxy. 
As a matter of fact, both early and late-type AGN2 host galaxies lie below the scaling relations defined by RM AGN1 and quiescent galaxies. 
Our analysis is rather conservative as we assumed that 
both RM AGN1 and AGN2 share the same average virial factor, 
$\langle f\rangle=4.31$. 
In order to have the same BH masses at fixed $\upsigma_\star \sim 185$ km s$^{-1}$,
the AGN2 should have a virial factor $\sim$7 times higher than 
the RM AGN1.
However, as also discussed in Paper II, according to the 
AGN unified model, AGN2 are viewed at larger inclinations (more edge-on) than AGN1, and 
there are indications that the $f$-factor 
decreases with increasing inclination \citep{risaliti11, pancoast14, bisogni17}. This would imply that 
an even smaller $f$-factor 
would probably be more appropriate for AGN2.
This argument is also supported by the recent finding 
of \citet{du17} who performed a spectropolarimetric study of 
six Seyfert 2 having dynamically measured $M_{\rm BH}$, and found 
a $f$-factor consistent with that of pseudo-bulges.  

At face value, our sample of AGN2 do not follow the 
scaling relation determined for RM AGN1 \citep{woo13}, 
as they seem to follow a shallower relation with $\upsigma_\star$.
Indeed this can be also seen in the 
bottom panel of Figure \ref{Fig:Msbin}, where the 
residuals from the scaling relation have a dependence on 
$\upsigma_\star$. 
The AGN2 lie below all the scaling relations presented in Figure \ref{Fig:Msbin}, which have been calibrated on 
quiescent galaxies \citep{woo13,kormendyho13,hk14,savorgnan15}. 
In particular, the scaling relations of \citet{kormendyho13}
is valid for elliptical and classical bulges, whereas the 
relation of \citet{hk14} is 
calibrated on pseudo-bulges. 
However, the determination of 
a different $M_{\rm{BH}}-\upsigma_\star$
relation for AGN2 is beyond the scope of this letter.

Recently, \citet[][triple-dot-dashed line in Figure \ref{Fig:Msbin}]{shankar16} discussed how 
the presence of selection effects in favour of 
the more massive BHs could be 
important in determining the 
underlying $M_{\rm{BH}}-\upsigma_\star$ relation. 
In this framework our AGN2 dataset, where faint broad virialized emission lines have been detected, 
suggests that the whole AGN population could indeed
follow a scaling relation with a lower normalization 
and a broader spread
than previously measured.  
As more dynamically measured $M_{\rm BH}$ are collected also for low-massive BHs, $M_{\rm BH}<10^7$ M$_\odot$, the full distribution of BH masses at fixed galaxy properties is now starting to be explored.
Indeed our results on the AGN2 
$M_{\rm BH} -\upsigma_\star$ are also in 
agreement with the $M_{\rm BH} -\upsigma_\star$ 
relation measured in megamaser disks galaxies \citep{greene16}. 
As discussed in \citet{reines15}, 
there is the possibility that if BH seeds are 
massive (e.g. $M_{\rm BH}=10^5$ M$_\odot$) 
the low-mass end of the relation between BHs and galaxies flattens toward an asymptotic value, creating a characteristic ``plume'' of less grown BHs \citep[see also][]{barausse17}.

The observed difference in BH masses in type 1 and 2 AGN sharing the same luminosity (Paper II) and velocity dispersion
does not comply with the standard AGN unified model
where the type 1 and 2 classification is only the product of line-of-sight orientation. 
Also the modified AGN unification scenario 
in which the torus inner radius (and then the opening angle) 
increases with the 
AGN luminosity \citep{lawrence91} is not able to reproduce our observations, 
as we see a difference in BH masses at fixed intrinsic AGN luminosity and host galaxy properties. 
Evolutionary models \citep{hopkins05} that predict AGN2 as 
the preceding buried accreting phase of an AGN1 
are instead favoured by our results.
However, given our current understanding of the growth of BHs in low- and moderate-luminosity AGN 
\citep[and in particular the importance of stochastic 
accretion and variability, e.g.][]{hickox14,schawinsk15}, this explanation 
could probably be 
incomplete.
If the BH growth in type 1 and 2 AGN is driven mainly by galaxy mergers, it will result in different bulge properties than if the evolution is mainly driven by internal secular processes. 
This issue will be further discussed in a forthcoming paper (Ricci et al. in prep).

As the BH-host galaxy scaling relations are 
the fundamental ingredients to derive the BHMF and 
its evolution, it is mandatory to better 
quantify the observed BH mass differences in type 1 and 2 AGN.
As discussed in the introduction, the 
scaling relations could be different according to 
the bulge host morphology. This translates into 
different $f$-factors and potentially different $M_{\rm{BH}}$.
It is thus important to analyse the 
AGN host bulge morphology to better 
understand which are the main drivers of the 
observed differences between type 1 and 2 AGN.

\section*{Acknowledgments}
\footnotesize	
	Part of this work was supported by PRIN/MIUR 2010NHBSBE and PRIN/INAF 2014\_3. 
	We thank the referee for constructive comments which improved the quality of our paper.
	We thank Cesare Perola for useful discussions.
	RS acknowledges support from the ERC Grant Agreement n. 306476.
	RM acknowledges ERC Advanced Grant 695671 ''QUENCH'' and
	support by the Science and Technology Facilities Council (STFC). 
	This paper is based on observations made with ESO telescopes at the Paranal Observatory and the Large Binocular Telescope (LBT) at Mt. Graham, Arizona.
	This research has made use of 
	the HyperLeda database and of
	the NASA/IPAC Extragalactic Database (NED), which is operated by the Jet Propulsion Laboratory, California Institute of Technology, under contract with the National Aeronautics and Space Administration.
\normalsize

\bibliographystyle{mnras} 
\bibliography{mybib} 

\label{lastpage}

\end{document}